\documentclass[]{article}

\usepackage{amsmath}
\usepackage{amssymb}
\usepackage{geometry}
\usepackage{xcolor}
\usepackage{multirow}
\usepackage{makecell}
\usepackage{hyperref}
\usepackage{lineno}
\usepackage{authblk}
\usepackage{graphicx}

\title{Searches for neutrinos in the direction of radio-bright blazars with the ANTARES telescope}

\date{}

\author[ ]{ANTARES Collaboration:}
\author[1,2]{A.~Albert}
\author[3]{S.~Alves}
\author[4]{M.~Andr\'e}
\author[5]{M.~Ardid}
\author[5]{S.~Ardid}
\author[6]{J.-J.~Aubert}
\author[7]{J.~Aublin\thanks{Corresponding author}}
\author[7]{B.~Baret}
\author[8]{S.~Basa}
\author[7]{Y.~Becherini}
% ...

\affil[1]{\scriptsize{Universit\'e de Strasbourg, CNRS,  IPHC UMR 7178, F-67000 Strasbourg, France}}
\affil[2]{\scriptsize{Universit\'e de Haute Alsace, F-68100 Mulhouse, France}}
\affil[3]{\scriptsize{IFIC - Instituto de F\'isica Corpuscular (CSIC - Universitat de Val\`encia) c/ Catedr\'atico Jos\'e Beltr\'an, 2 E-46980 Paterna, Valencia, Spain}}
\affil[4]{\scriptsize{Technical University of Catalonia, Laboratory of Applied Bioacoustics, Rambla Exposici\'o, 08800 Vilanova i la Geltr\'u, Barcelona, Spain}}
\affil[5]{\scriptsize{Institut d'Investigaci\'o per a la Gesti\'o Integrada de les Zones Costaneres (IGIC) - Universitat Polit\`ecnica de Val\`encia. C/  Paranimf 1, 46730 Gandia, Spain}}
\affil[6]{\scriptsize{Aix Marseille Univ, CNRS/IN2P3, CPPM, Marseille, France}}
\affil[7]{\scriptsize{Universit\'e Paris Cit\'e, CNRS, Astroparticule et Cosmologie, F-75013 Paris, France}}
\affil[8]{\scriptsize{Aix Marseille Univ, CNRS, CNES, LAM, Marseille, France }}
% ...

\author[ ]{\ \\ \ \\}
\author[45,a]{Y.~A.~Kovalev}
\author[46,45,47,a]{Y.~Y.~Kovalev}
\author[48,45,a]{A.~V.~Plavin\thanks{Corresponding author}}
\author[49,45,a]{A.~B.~Pushkarev}
\author[50,51,a]{S.~V.~Troitsky}

\affil[45]{\scriptsize{Lebedev Physical Institute of the Russian Academy of Sciences, Leninsky prospekt 53, 119991 Moscow, Russia}}
\affil[46]{\scriptsize{Max-Planck-Institut f\"ur Radioastronomie, Auf dem H\"ugel 69, 53121 Bonn, Germany}}
\affil[47]{\scriptsize{Moscow Institute of Physics and Technology, Institutsky per. 9, Dolgoprudny 141700, Russia}}
\affil[48]{\scriptsize{Black Hole Initiative at Harvard University, 20 Garden Street, Cambridge, MA 02138, USA}}
\affil[49]{\scriptsize{Crimean Astrophysical Observatory, 298409 Nauchny, Crimea, Russia}}
\affil[50]{\scriptsize{Institute for Nuclear Research of the Russian Academy of Sciences, 60th October Anniversary Prospect 7a, Moscow, Russia}}
\affil[51]{\scriptsize{Faculty of Physics, M.V. Lomonosov Moscow State University, 1-2 Leninskie Gory,  Moscow 119991, Russia}}

\author[ ]{\ \\ \ \\}
\author[ ]{OVRO Collaboration:}
\author[52,53]{T. Hovatta}
\author[54,55]{S. Kiehlmann}
\author[52]{I. Liodakis}
\author[54,55]{V. Pavlidou} 
\author[56]{A.C.S Readhead}

\affil[52]{\scriptsize{Finnish Center for Astronomy with ESO, University of Turku, Vesilinnantie 5, FI-20014, Finland}}
\affil[53]{\scriptsize{Aalto University Mets\"ahovi Radio Observatory, Mets\"ahovintie 114, FI-02540 Kylm\"al\"a, Finland}}
\affil[54]{\scriptsize{Institute of Astrophysics, Foundation for Research and Technology-Hellas, GR-70013 Heraklion, Greece}}
\affil[55]{\scriptsize{Department of Physics and Institute of Theoretical and Computational Physics, University of Crete, 70013 Heraklion, Greece}}
\affil[56]{\scriptsize{Owens Valley Radio Observatory, California Institute of Technology,  Pasadena, CA 91125, USA}}

\affil[a]{\scriptsize{The shown author affiliations reflect their job contracts; the ANTARES collaboration has currently suspended all institutional relations with Russian science organisations.}}

\begin{document}

\maketitle

\begin{abstract}

Active galaxies, especially blazars, are among the most promising neutrino source candidates.
To date, ANTARES searches for these objects considered GeV-TeV $\gamma$-ray bright blazars.
Here, a statistically complete radio-bright blazar sample is used as the target for searches of origins of neutrinos collected by the ANTARES neutrino telescope over 13 years of operation.
The hypothesis of a neutrino-blazar directional correlation is tested by pair counting and by a complementary likelihood-based approach.
The resulting post-trial $p$-value is $3.0\%$ ($2.2\sigma$ in the two-sided convention), possibly indicating a correlation. 
Additionally, a time-dependent analysis is performed to search for temporal clustering of neutrino candidates as a mean of detecting neutrino flares in blazars.
None of the investigated sources alone reaches a significant flare detection level. However, the presence of 18 sources with a pre-trial significance above $3\sigma$ indicates a $p=1.4\%$ ($2.5\sigma$ in the two-sided convention) detection of a time-variable neutrino flux. An \textit{a posteriori} investigation reveals an intriguing temporal coincidence of neutrino, radio, and $\gamma$-ray flares of the J0242+1101 blazar at a $p=0.5\%$ ($2.9\sigma$ in the two-sided convention) level. Altogether, the results presented here suggest a possible connection of neutrino candidates detected by the ANTARES telescope with radio-bright blazars.
\end{abstract}

\section{Introduction}
\label{sec:intro}

Astrophysical neutrinos of high energies were discovered almost ten years ago by the IceCube Collaboration  \cite{IceCube:2014stg}, consistently with later studies with the other two existing instruments, namely ANTARES \cite{Fusco:2020owb} and Baikal-GVD \cite{Baikal-GVD:2022fis}. However, their origin is still not fully determined. Observational evidence exists for a wide range of neutrino sources, including individual active galaxies \cite{IceCube:2018cha, IceCube:2018dnn, IceCube:2022der}, large samples of blazars \cite{Plavin:2020emb, Plavin:2020mkf, Plavin:2022oyy, Kun:2022khf, Buson:2022fyf, Giommi:2020hbx, IceCubeCollaboration:2022fxl, Buson:2023irp}, and the Milky Way Galaxy \cite{ANTARES:2017nlh, ANTARES:2018nyb, ANTARES:2022izu, Neronov:2015osa, IceCube:2019lzm, IceCube:2023ame, Kovalev:2022izi}.

Active galaxies, and particularly blazars, which have their jets pointed towards us, form a very promising class of neutrino source candidates \cite{BerezinskyNeutrino77,Eichler1979,Murase:2015ndr, Boettcher:2019gft}. The emission from the jet is detectable across cosmological distances thanks to the relativistic beaming \cite{2019ARA&A..57..467B}. Similar beaming effects should affect neutrinos as well, enhancing their chance to be detected on Earth. Moreover, there are hints that neutrinos are predominantly produced during major flares close to the jet origin \cite{Plavin:2020emb, Hovatta:2020lor}, indicating a tight physical connection between jets and neutrino production processes.

Currently, all associations of high-energy neutrinos with celestial objects or classes thereof are fundamentally statistical. The comparison of findings based on different instruments is crucial to understand both astrophysical neutrino flux characteristics, such as its energy spectrum, and potential systematic effects inherent to their detection. Earlier searches for a neutrino signal in the ANTARES detector from the direction of blazars were performed using $\gamma$-bright blazars contained in the \textit{Fermi}~LAT catalog \cite{ANTARES:2015gxt, ANTARES:2020zng}. However, recent studies suggest that emission and flares in $\gamma$-rays may not be tightly correlated with neutrino sources \cite{IceCube:2020wum, Plavin:2020mkf, Boettcher:2022dci}: hadronic $\gamma$-rays are co-produced together with neutrinos, but quickly cascade down in energy. Synchrotron emission from blazar jets, detected on Earth as radio emission, could likely be a better tracer of relativistic beaming and activity happening close to the jet origin. 

In addition, the recent publication of the IceCube Event Catalog of Alert Tracks-1 \cite{IceCube:2023agq}, containing a revised sample of neutrino candidates with a high probability of being of astrophysical origin, has been followed by a search~\cite{IceCube:2023htm} for a correlation with the population of blazars contained in the \textit{Fermi} 4LAC-DR2 catalog \cite{Fermi-LAT:2019pir} and in the Radio Fundamental Catalog\footnote{\url{http://astrogeo.smce.nasa.gov/rfc/}}. As no significant correlation is found, this result mitigates the previous findings of \cite{Plavin:2020emb}, and indicates that only a minority of the $\gamma$-ray bright and radio-bright AGN population can be identified as potential sources of high-energy neutrinos.

This paper presents an independent and complementary study of the possible neutrino-blazar association with data collected by the ANTARES observatory. Section~\ref{sec:sample} and Section~\ref{sec:vlbi} define the dataset used in the analysis: the ANTARES neutrino candidates and the radio observations of inner blazar regions. A description of the employed analysis methods and respective results follows. The hypothesis of directional correlation between neutrinos and blazar emission is tested by mean of a neutrino-blazar pair counting method (Section~\ref{sec:counting}), as well as of a complementary time-integrated likelihood-based approach (Section~\ref{sec:time-int-lik}). Moreover, in order to search for a time clustering of ANTARES events coming from the blazar directions, a time-dependent likelihood scan is also performed (Section~\ref{sec:flares}). In Section~\ref{sec:MMflares}, the results of a follow-up search for multi-messenger time flare associations are presented. Finally, Section~\ref{sec:conclusions} summarizes all findings and discusses further prospects.

\section{ANTARES detector and data sample} 
\label{sec:sample}

ANTARES~\cite{antaresdetector} was an undersea high-energy neutrino telescope, 
located 40~km off-shore from Toulon, France, below the surface of the Mediterranean Sea. After over 15 years of operation, it took its last data in February 2022. The detector consisted of a three-dimensional array of 885~photomultiplier tubes (PMTs), distributed along 12 vertical lines of 450~m length, for a total instrumented volume of $\sim$ 0.01$\ \textrm{km}^3$. Each line was anchored to the seabed at a depth of about 2500~m and held taut by a buoy at the top.
The PMTs collected the Cherenkov photons induced by the passage of the relativistic charged particles produced in neutrino interactions inside or near the instrumented volume. The information provided by the position, time and collected charge of the signals in the PMTs can be used to infer the direction and energy of the incident neutrino. 

Two main event topologies, induced by different neutrino flavours and types of interactions, could be identified: track-like and shower-like events. Charged current (CC) interactions of muon neutrinos produce relativistic muons that can travel large distances through the medium, with Cherenkov light being induced all along the muon path, resulting in a track-like signature in the detector. The parent neutrino direction of high-quality selected tracks can be reconstructed with a median angular resolution of~$\sim$ 0.8$^{\circ}$ at~$E_{\nu}$~$ \sim$ 1~TeV and below~$\sim$ 0.4$^{\circ}$ for $E_{\nu} > 10$~TeV~\cite{ANTARES:2017dda} thanks to the long lever arm of this channel. 
Shower-like events are induced by all-flavour neutral current (NC) as well as ${\nu_e}$ and ${\nu_\tau}$~CC interactions. This topology is 
characterised by an almost spherical light emission around the shower maximum, with an elongation of a few metres, which results in a worse estimation of the parent neutrino direction compared to the track channel. A median angular resolution of~$\sim$ 3$^{\circ}$ is achieved for high-quality selected showers with energies between 1~TeV and 0.5~PeV~\cite{ANTARES:2017dda}.

The data set employed in this analysis includes events recorded in ANTARES between January 29, 2007 and February 29, 2020 (3845 days of livetime), and selected using the criteria defined in~\cite{ANTARES:2017dda}, optimized to minimise the neutrino flux needed for a $5\sigma$ discovery of a point-like source 
with a $\propto E^{-2.0}$ emission spectrum.
The selection includes cuts on the zenith angle, the angular error estimate and parameters describing the quality of the reconstruction. In the shower channel, an additional cut is applied on the interaction vertex, required to be located within a fiducial volume slightly larger than the instrumented volume. 
Recently improved calibrations have been used to reconstruct all the selected ANTARES events. This yields slightly different values of the reconstructed direction -- with $\sim$ $98\%$ of the selected events being reconstructed within $1^{\circ}$ from the previous direction --
and of the quality parameters associated with each event. 
In addition, the estimator of the energy for track-like events has been updated, taking into account the time-evolution of the detector over the whole data-taking period. 
A total of 10504 track-like and 227 shower-like events survive the selection, with an expected atmospheric muon contamination of $\sim$ $14\%$ and $\sim$ $43\%$ for the track and the shower channel, respectively.  Only the track channel is employed in the search for directional correlations, while the search for neutrino temporal flares makes use of both tracks and showers.

\section{Radio observations of blazars}
\label{sec:vlbi}

Synchrotron emission from inner regions of blazar jets is observed and measured by radio telescopes using very long baseline interferometry (VLBI) techniques. VLBI-measured flux densities represent the emission from the very central parsecs of blazars, and these measurements are used to select the objects included in this analysis.

\begin{figure}[h]
    \centering
    \includegraphics[width=0.9\linewidth]{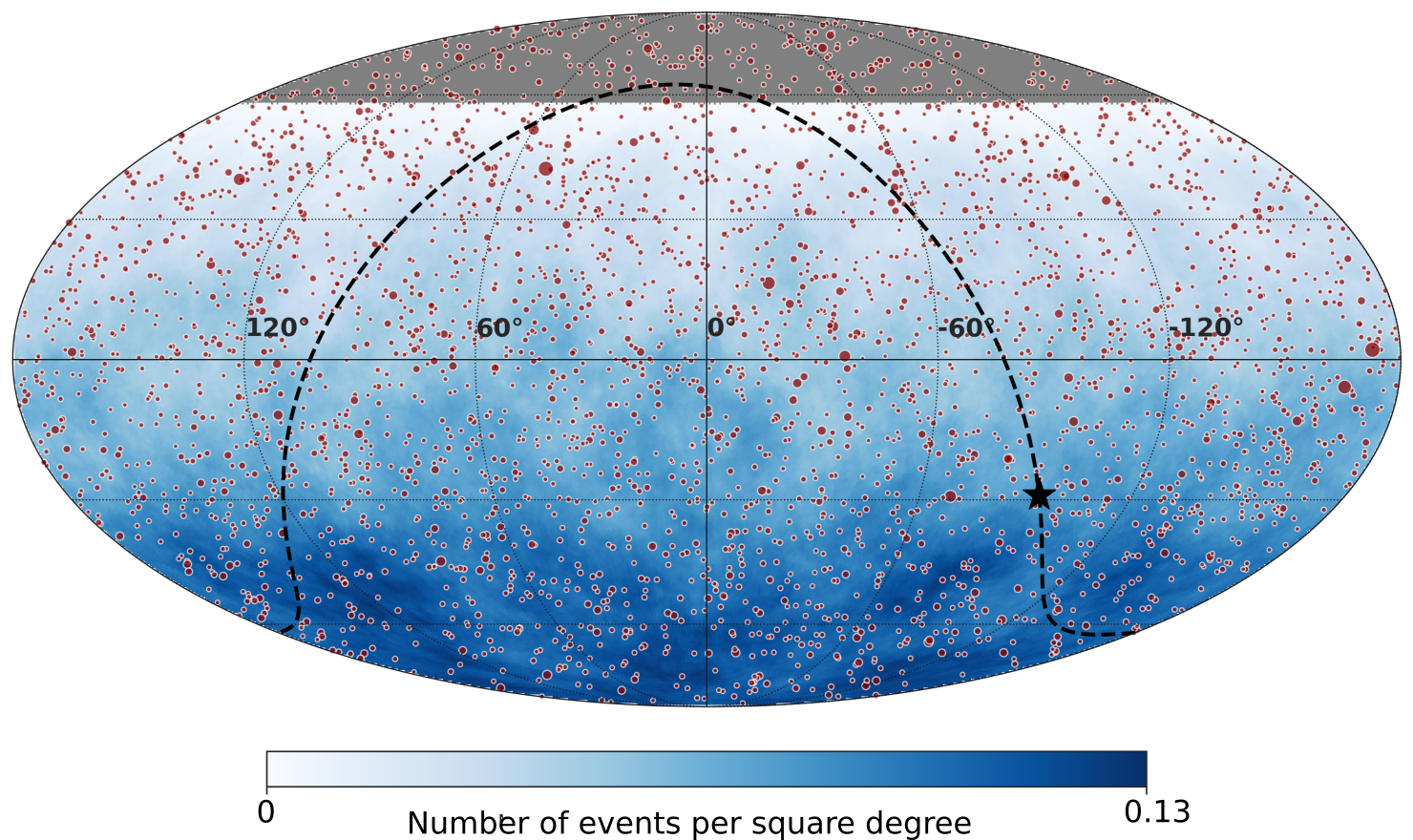}
    \caption{Equatorial sky map showing the location of the VLBI blazars (red dots), together with the average density of ANTARES track-like and shower-like events per square degree. The event density at a given point in the sky is obtained dividing the number of events found within a cone of $10^\circ$ radius around this point by the solid angle of the cone. The surface of the markers is proportional to the VLBI $8$ GHz flux density. The blazars located in the grey region at high declinations are outside the ANTARES field of view. The dashed black line shows the Galactic plane, and the Galactic center is represented as a black star marker.
    }
    \label{fig:SkymapFull}
\end{figure}

A complete flux-limited sample of blazars observed by VLBI is selected using the same criteria as in \cite{Plavin:2020mkf}. Accurate positions and parsec-scale flux densities of observed blazars are compiled in the Radio Fundamental Catalogue\footnote{\url{http://astrogeo.smce.nasa.gov/sol/rfc/rfc\_2020d/}}. Observations at 8~GHz are utilized because of their completeness across the whole sky. There are 3411~blazars with historical average flux density above 150~mJy; 3051 of them fall into the ANTARES field of view and are targeted by the searches for directional correlations. Figure \ref{fig:SkymapFull} shows the location of the selected blazars in equatorial coordinates, together with the ANTARES neutrino candidates. As for the time-dependent scan, the sources with low visibility, i.e.\ located above $40$$^{\circ}$ in declination, are excluded from the search, leading to a targeted catalog of 2774 blazars.
The collection of observations includes geodetic VLBI \cite{2009JGeod..83..859P,2012A&A...544A..34P,2012ApJ...758...84P}, the Very Long Baseline Array (VLBA) calibrator surveys (VCS \cite{Beasley:2002yf,2003AJ....126.2562F, Petrov:2004qm, Petrov:2005fd, Kovalev:2006td, Petrov:2008nn, r:wfcs, 2016AJ....151..154G}), other 8~GHz global VLBI, VLBA, EVN (the European VLBI Network) and LBA (the Australian Long Baseline Array) programs 
\cite{2011AJ....142...35P, 2011AJ....142..105P, 2011MNRAS.414.2528P, 2012MNRAS.419.1097P, Petrov:2013sc, Schinzel:2014eja, 2017ApJS..230...13S, 
2019MNRAS.485...88P}.

\section{Counting method}\label{sec:counting}
The first correlation analysis uses a simple method, inspired by \cite{Plavin:2020emb}, where the observable is the number of neutrino-blazar angular pairs separated by an angular distance $\Psi$ less than $x \cdot \beta$. Here, $\beta$ is the angular uncertainty coming from the neutrino reconstruction, and $x$ is a free parameter that varies in the interval $[0.1;2]$. The parameter $x$ is meant to take into account a possible systematic difference between the output of the reconstruction algorithm and the true (unknown) angular error radius. Using Monte Carlo simulations, the relationship between the error estimate $\beta$ and the true $68\%$ containment radius $\Psi_{68\%}$ can be assessed. For values $\beta \lesssim 0.5$, and reconstructed energies above $\sim 10$ TeV, the relation between $\beta$ and $\Psi_{68\%}$ is close to what is expected from a two-dimensional Gaussian function. At lower energies and for higher values of $\beta$, the value of $\Psi_{68\%}$ becomes significantly higher than the Gaussian expectation. The scan on the $x$ parameter is then an empirical way to take into account this complicated behavior while still using the reconstruction quality information of ANTARES neutrinos on an event-by-event basis. As a consequence, the $p$-value obtained with this method needs a trial factor correction.

\begin{figure}[]
    \centering
    \includegraphics[width=0.9\linewidth]{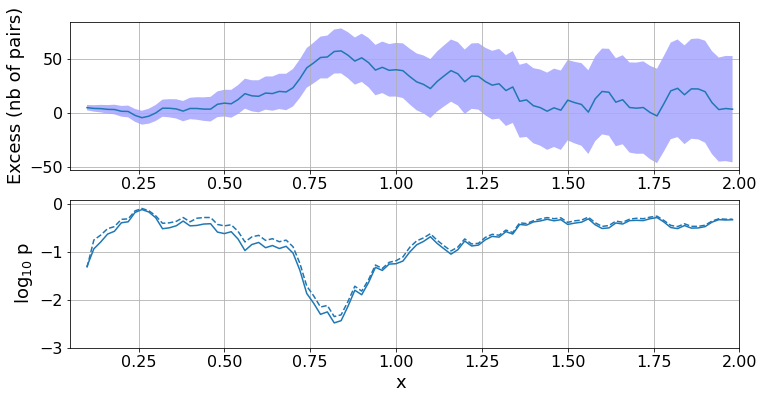} 
    \caption{Result of the counting analysis with the VLBI blazar catalog. The top and bottom panels show, as a function of the parameter $x$ (defined in the text), the observed excess of pairs relative to random expectations, and the pre-trial $p$-value respectively. In the top panel, the blue band shows the $\pm 1 \, \sigma$ confidence interval. In the bottom panel, the dashed curve shows the $p$-value obtained when excluding blazars with $S>3.68\, \mathrm{Jy}$ from the catalog.}
    \label{fig:1Dscan}
\end{figure}

A possible correlation between the radio-flux density and neutrino emission is evaluated by performing an additional scan on the radio-flux density $S_{8\mathrm{GHz}}$: VLBI blazars are kept in the sample if they satisfy $S_{8\mathrm{GHz}} > S_\mathrm{min}$, and the value of $S_\mathrm{min}$ is varied in the range $[0.15;5]$ Jy.

\begin{figure}[h!]
    \centering
    \includegraphics[width=0.8\linewidth]{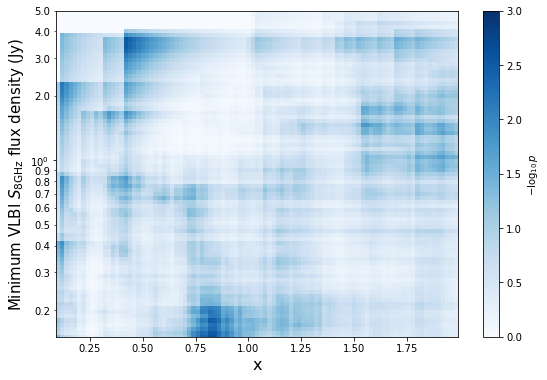}
    \caption{Result of the two-dimensional scan over the radio-flux density $S_{8\mathrm{GHz}}$ and parameter $x$. The color code indicates the pre-trial $p$-value.
    }
    \label{fig:2Dscan}
\end{figure}

The results of the counting analysis are illustrated in Figures~\ref{fig:1Dscan} and~\ref{fig:2Dscan}. 
When using the full VLBI catalog, the one-dimensional scan shows an absolute minimum for $x=0.82$, where $n_\mathrm{obs}=469$ pairs are observed in data, while $n_\mathrm{exp}=410.4$ are expected on average from random simulations (59 pairs in excess). The associated pre-trial $p$-value is $p=2.5 \times 10^{-3} \, (3.0 \sigma)$, leading after correction to a post-trial $p$-value of $P=3.0 \times 10^{-2} \; (2.2\sigma)$.
Note that without performing a scan, the value $p(x=1) = 5.0 \times 10^{-2}$ $(2.0 \sigma)$ would have been quoted instead.

The position $x=0.82$ of the minimum seems to indicate that the angular uncertainty $\beta$ of the neutrinos is over-estimated, which is not what is expected from Monte Carlo. However, the position of this scan minimum is subject to random fluctuations, and as the excess is not statistically significant, this observation cannot be interpreted as evidence for an issue in the ANTARES track reconstruction.

The search for a correlation with radio flux density with the two-dimensional scan ($x,S_\mathrm{min}$) is shown in Figure \ref{fig:2Dscan}. The absolute minimum is found for $x=0.82$ and $S_\mathrm{min}=0.15$ Jy, with a pre-trial $p$-value of $p (S>0.15\, \mathrm{Jy})=2.5 \times 10^{-3}$, and a post-trial $P(S>0.15\, \mathrm{Jy})=0.26$. This minimum corresponds to the previous findings of the one dimensional scan, and is obtained for the lowest value of the flux density cut, meaning that the whole VLBI blazar catalog is included. 

A local minimum is also visible for $x=0.42$ and $S_\mathrm{min}=3.68$ Jy, with a pre-trial $p$-value of $p(S>3.68\, \mathrm{Jy})=2.7 \times 10^{-3}$. This excess is mainly driven by 3 blazars: J0609$-$1542 and J1743$-$0350, that have one very close ANTARES track (less than $0.2^\circ$ away), and J0538$-$4405 that has two events at $0.4^\circ$ distance. These sources are not found in the search for neutrino flares presented in Section \ref{sec:flares}, as only one neutrino is contributing for J0609$-$1542 and J1743$-$0350, and the two events located around J0538$-$4405 are separated by more than $6.5$ years in time.

When accounting for the trial factors, the significance of this excess is $P(S>3.68\, \mathrm{Jy})=0.28$. In addition, as can be seen in Figure~\ref{fig:1Dscan} (dashed line), when excluding the blazars with $S>3.68\, \mathrm{Jy}$ from the catalog, the position of the minimum stays the same, and the pre-trial $p$-value is only slightly increased $p=4.5 \times 10^{-3}$. These results indicate that the excess in the counting of neutrinos-blazar pairs is not induced by a small number of very high flux sources in the VLBI catalog.

\section{Time-integrated likelihood analysis}\label{sec:time-int-lik}

A time-integrated likelihood analysis very similar to the one reported in \cite{ANTARES:2020zng} is performed, making use of more information about the ANTARES detector response than the one used in the simple counting method.  The likelihood of the null hypothesis $H_b$ where only background is present is compared with an alternative hypothesis $H_{s+b}$ where signal events coming from blazars are present in the data.

The neutrino signal events are assumed to come from blazars in proportion to their measured VLBI flux density $S_{8\mathrm{GHz}}$, with an energy spectrum modeled as a pure power-law $E^{-\gamma}$. The analysis is performed for a fixed value of the spectral index $\gamma$, and repeated for values $\gamma \in [1.8;2.6]$ in $0.1$ steps.

The likelihood is written as:

\begin{equation}  \label{eq:time_int_lik}
    \ln \mathcal L_\mathrm{s+b} = \sum_i^N \ln \left( \mu_s S_i + \mu_b B_i \right) - \mu_s - \mu_b ,
\end{equation} 

\noindent
where $N$ is the total number of observed track-like events, $S_i$ is the probability density function (PDF) of the signal and $B_i$ the background one. 
The free parameters are the estimated number of signal $\mu_s$ and background $\mu_b$ events. The expression of the background-only likelihood $ \mathcal L_\mathrm{b}$ is simply obtained by setting $\mu_s=0$ in Equation~\ref{eq:time_int_lik}.

The test statistics $Q$ is defined as a likelihood ratio:
\begin{equation}
    Q = 2 \ln \left[ \frac{\mathrm{max}(\mathcal L_\mathrm{s+b})}{ \mathrm{max}(\mathcal L_\mathrm{b})} \right],
    \label{eq:ts_ti}
\end{equation}
where the likelihoods defined in Equation~\ref{eq:time_int_lik} are maximized with respect to the free parameters $\mu_s$ and $\mu_b$. In practice, as the background-only term $\mathcal L_\mathrm{b}$ is maximal when $\mu_b=N$, the numerical procedure of likelihood maximization has to be performed only for the $\mathcal L_\mathrm{s+b}$ term.\\

The core ingredients of the likelihood analysis are the signal and background PDF, which are written as the product of a spatial and an energy term:
\begin{equation}
 S_i  =f_s(\alpha_i,\delta_i,\beta_i,E_i) \, g_s(E_i) \quad \mathrm{and} \quad 
  B_i  =f_b(\delta_i,E_i) \, g_b(E_i),  
   \label{eq:PDFs}
\end{equation}
where $(\alpha_i,\delta_i)$ are the equatorial coordinates, $\beta_i$ the angular uncertainty and $E_i$ the estimated energy of the $i^{th}$ neutrino candidate. 

The energy-dependent part of the signal PDF, $g_s(E)$, is computed using Monte Carlo simulations, by building a histogram of the reconstructed energies $E$ of track-like events when assuming a pure power-law $E_\mathrm{true}^{-\gamma}$ distribution of the true neutrino energies $E_\mathrm{true}$. The background PDF, $g_b(E)$, is similarly obtained by building a histogram with Monte Carlo simulations, where the  atmospheric neutrino fluxes (conventional+prompt components) and atmospheric muons fluxes are taken into account (see \cite{ANTARES:2017srd} for details). \\

The spatial term $f_b(\delta_i,E_i)$ for the background is considered to be independent of the right ascension $\alpha$. Indeed, the ANTARES data set has been obtained by accumulating fifteen years of quasi-continuous measurement, therefore the non-uniformity of the detector exposure in local coordinates is averaged out by the rotation of the Earth, leading to a flat right ascension distribution of events. In practice, the value of the background spatial term $f_b(\delta_i,E_i)$ is estimated by a linear interpolation within a two-dimensional ($\sin \delta, E)$ histogram built from Monte Carlo simulations. 

The spatial signal term is obtained by summing over all the individual blazars contributions: 
\begin{equation}
   f_s(\alpha_i,\delta_i,\beta_i,E_i)= \frac{1}{\sum w_j} \sum_{j=1}^{N_\mathrm{sources}} w_j \, \mathcal{F}(\Psi_{ij},\,\delta_i, \,\beta_i\,,E_i) \quad \quad w_j= w_j^\mathrm{model} \, \mathcal{A}(\delta_j),
   \label{eq:signalPDF}
\end{equation}
where $\mathcal{F}(\Psi_{ij},\,\delta_i, \,\beta_i\,,E_i)$ is the ANTARES point spread function (PSF) for track-like events, defined as the probability density for the event direction to fall within a given angular distance from the true incoming direction. The PSF is a sharply decreasing function of the space angle $\Psi_{ij}$ between the $i^\mathrm{th}$ neutrino and the $j^\mathrm{th}$ blazar, and mostly depends, by decreasing order of importance, on the angular error $\beta$, the reconstructed energy $E$, and the declination $\delta$. 
The practical implementation of the PSF is obtained from Monte Carlo simulations, by building two-dimensional histograms in $(E, \beta)$ for $8$ separate bins of $\sin \delta$.

The weight of the $j^\mathrm{th}$ blazar is proportional to its measured flux density $w_j^\mathrm{model}=S_{8\mathrm{GHz}}$, and corrected by the declination-dependent acceptance $\mathcal{A}(\delta)$ of the ANTARES neutrino track sample (computed for the particular $E^{-\gamma}$ energy spectrum considered). For comparison, a basic scenario where all blazars have the same weight, $w_j^\mathrm{model}=1$, is considered.

The smallest $p$-value obtained with the time-integrated likelihood analysis is $p=2.6 \times 10^{-2}$ $(2.2\sigma)$, for a $E^{-2.3}$ neutrino energy spectrum and with the radio-flux weight hypothesis. This spectral index value corresponds to the best-fit one from the ANTARES diffuse analysis \cite{Fusco:2020owb}, providing a hint that the VLBI blazars could contribute to the astrophysical neutrino diffuse flux. In comparison, when assuming an equal weight for each blazar in the likelihood, the resulting $p$-values ranges from $p=7.6 \times 10^{-2}$ $(1.8  \sigma)$ for $E^{-2.3}$ up to $p=0.38$ $(0.9 \sigma)$ for $E^{-1.8}$. The fact that smaller $p$-values are obtained with the flux-weight hypothesis could be an indication that the neutrino candidates correlate with blazars having a high flux density.
 
\begin{figure}[h!]
    \centering
    \includegraphics[width=0.9\linewidth]{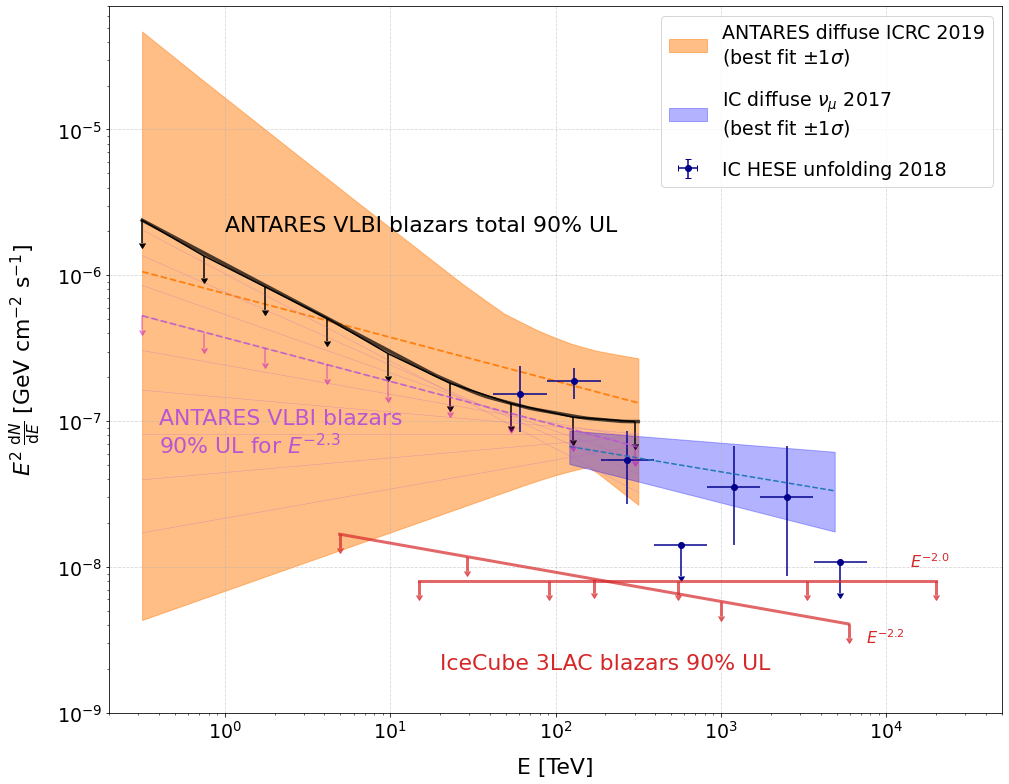} 
    \caption{Upper limits on the one-flavour $(\nu_\mu + \bar{\nu}_{\mu})$ total neutrino flux from the VLBI blazars obtained with the time-integrated likelihood analysis as a function of the neutrino energy. The ANTARES limits for each of the spectral indices tested are represented by thin solid violet lines. For comparison, the IceCube limits~\cite{ICBlazars_ICRC19} on \textit{Fermi} 3LAC blazars for $E^{-2.0}$ and $E^{-2.2}$ spectra are shown in red. The thick black line shows the highest values of the single spectral index limits, and provides the most conservative upper-limit curve. For comparison, the orange-shaded band shows the $68\%$ confidence region for the diffuse flux measured by ANTARES \cite{Fusco:2020owb}, the IceCube best fit to the astrophysical diffuse muon neutrino flux from \cite{Haack2017} is displayed as a blue-shaded band, and the IceCube HESE spectrum from \cite{Kopper2017} is shown as blue markers.
}
    \label{fig:LL_UL}
\end{figure}

As the studied spatial correlation between neutrino candidates and the VLBI blazars is not statistically significant, upper limits (UL) at $90\%$ confidence level are reported in Figure~\ref{fig:LL_UL}. The upper limits are computed for the different spectral indices of the assumed power-law neutrino flux (thin magenta lines), and their highest values as a function of the energy provides the most conservative limit curve (black dashed curve in the Figure). For comparison, the $68\%$ allowed values from the ANTARES diffuse analysis are plotted (with a $15\%$ systematic on the flux normalization included as in \cite{ANTARES:2017srd}), together with results from IceCube \cite{ICBlazars_ICRC19}.

As mentioned at the beginning of this section, our best fit for a cosmic diffuse neutrino flux (orange dotted line) has a spectral index of $\gamma=2.3$, and can be compared to the corresponding blazar upper limit (dotted magenta line). For this particular value of $\gamma$, the ratio of the magenta and orange dotted lines is $\sim 0.2$, implying that the VLBI blazars could not contribute to more than $\sim 20 \%$ of our estimated total diffuse flux of cosmic neutrinos. However, as can be seen on Figure~\ref{fig:LL_UL}, when accounting for the $68\%$ confidence interval on the ANTARES diffuse flux estimation, the total VLBI upper-limit (black line) only weakly constrains our measurement.
The total neutrino upper limit mildly constrains the first data point (blue marker) above $100$~TeV of the IceCube HESE measurement, where the VLBI blazars would contribute $\sim 50\%$ of the flux.

\section{Time-dependent likelihood scan} 
\label{sec:flares}

The time-dependent scan looks for neutrino flares from the direction of the selected radio-bright blazars and also relies on an unbinned maximum likelihood method. 
Since this search looks for clustering of events in space and time, the knowledge of the detection time of the selected events is included in the likelihood and combined with the spatial and energy information. This is achieved by multiplying the signal and background PDFs of Equation~\ref{eq:PDFs} by a time-dependent term. Unlike in the time-integrated analysis, the likelihood is maximised independently at the position of each investigated source, meaning that each source is analysed separately. Therefore, the spatial signal term of Equation~\ref{eq:signalPDF} simplifies into $f_s(\alpha_i,\delta_i,\beta_i,E_i)=\mathcal{F}(\Psi_{ij},\,\delta_i, \,\beta_i\,,E_i)$.
Regarding the signal time PDFs, two generic time profiles describing a temporary increase in neutrino emission -- a Gaussian profile and a box profile -- are tested. They are defined as: 

\begin{align}
\mathcal{S}_{\rm Gaussian}^{\rm time}(t_i) = \frac{1}{\sqrt{2\pi}\sigma_t}e^{-\frac{(t_i - T_{0})^2}{2\sigma_t^2}}, 
&&
\mathcal{S}_{\rm box}^{\rm time}(t_i)=\begin{cases}
    \frac{1}{2\sigma_{t}}, & \text{if $[T_{0} - \sigma_{t}] \leq t_i \leq [T_{0} + \sigma_{t}]$} ;\\
    0, & \text{otherwise} ;
  \end{cases} ,
\end{align}

\noindent with $t_i$ being the detection time of the neutrino candidate event $i$, and $T_{0}$ and $\sigma_t$ being the unknown central time and duration of the flaring emission, respectively, both fitted in the likelihood maximisation. Given the small expected contribution of a cosmic signal in the overall data set, the background time profile is built using the time distribution of data events, ensuring a time profile proportional to the measured data. This PDF is computed applying less stringent selection criteria than those of the final sample so as to avoid statistical fluctuations, using the same approach as in~\cite{ANTARES:2019itg}.

At the location of each investigated source, the likelihood is maximised leaving as free parameters the number of signal events $\mu_s$, the signal spectral index $\gamma$, the central time of the flare $T_0$, and the flare duration $\sigma_t$. The procedure allows for a single flare per blazar to be fitted, and, for the given best-fit flare (flare with highest $Q$), provides the best-fit values $\hat{\mu}_s$, $\hat{\gamma}$, $\hat{T}_0$, $\hat{\sigma}_t$.
In the maximisation, the spectral index can take values between 1.0 and 3.5. The range includes the value predicted by the candidate acceleration mechanism ($\gamma = 2.0$) and the softer best-fit spectral indices measured by the IceCube Collaboration for the diffuse neutrino flux ($\gamma = 2.37$~\cite{IceCube:2021uhz}, $\gamma = 2.53$~\cite{IceCube:2020acn}) and for the point-like sources showing evidence for neutrino emission, TXS~0506+056 ($\gamma = 2.1$ and $\gamma = 2.2$~\cite{IceCube:2018cha}) and NGC~1068 ($\gamma = 3.2$~\cite{IceCube:2022der}).
Concerning the time-dependent parameters, $T_0$ can vary over the time range of the investigated ANTARES data (from January 29, 2007 until February 29, 2020), while $\sigma_t$ can take values between $1$~day and $2000$~days. 

The test statistic $Q$, similarly to the time-integrated analysis, is defined as the logarithm of the ratio between the likelihood evaluated with the best-fit values of the free parameters and the background-only likelihood. However, in this case, the likelihood ratio is multiplied by a penalty term for short flares, $\frac{\hat{\sigma}_t}{\Delta T}$, with $\Delta T$ being the allowed time range for $T_0$. The purpose of the penalty term is to account for the larger trial factor that should be associated to short flares since a larger number of short flares than of long ones can be accommodated in a given time range, as described in~\cite{TimeDepMethod}.

In order to estimate the $p$-value of the best flare for each investigated source, the observed $Q$-value is compared to the test statistic distribution obtained with background-only pseudo-experiments (PEs). Each PE is obtained using sets of data randomised in right ascension to eliminate any local clustering due to potential sources keeping the corresponding source declination. In particular, the $p$-value is given by the fraction of background-like PEs with a value of the test statistic larger than the observed $Q$. The lowest obtained $p$-value identifies the most significant flare of the search. Finally, a trial correction that accounts for the fact that many candidates have been investigated is applied, by comparing the lowest obtained $p$-value to the distribution of the smallest $p$-values found when performing the same analysis on many background-only PEs.

\begin{figure*}[ht]
\centering
    \includegraphics[width=0.8\textwidth]{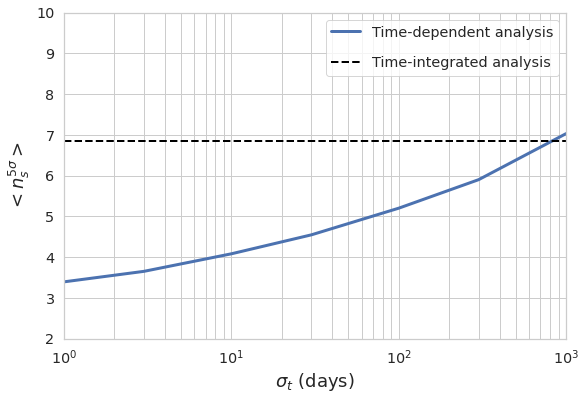} 
    \caption{$5\sigma$ discovery potential in terms of mean number of signal events as a function of the simulated flare duration for the time-integrated analysis (dashed) and for the time-dependent analysis (solid). The simulated source is at a declination of $\delta = -40^{\circ}$ and the flare is centred at $T_0$~[MJD]~$ = 57000$. Similar results are obtained for different source declinations and central times.}
    \label{fig:performance}
\end{figure*}

Figure~\ref{fig:performance} shows the expected performance of this approach compared to that of the time-integrated analysis, i.e.\ when the time information of the events is not considered. The $5\sigma$ discovery potential $<n_{s}^{5\sigma}>$ -- defined as the mean number of injected signal events needed for a 5$\sigma$ discovery in 50\% of the PEs -- is shown as a function of the duration of the flare. 
The time-dependent search performs better along almost all the investigated range of flare durations, with an improvement in the discovery potential by a factor~$\sim$ 2 achieved for flares as short as 1~day. 

\begin{figure}[h!]
\centering
    \includegraphics[width=0.8\textwidth]{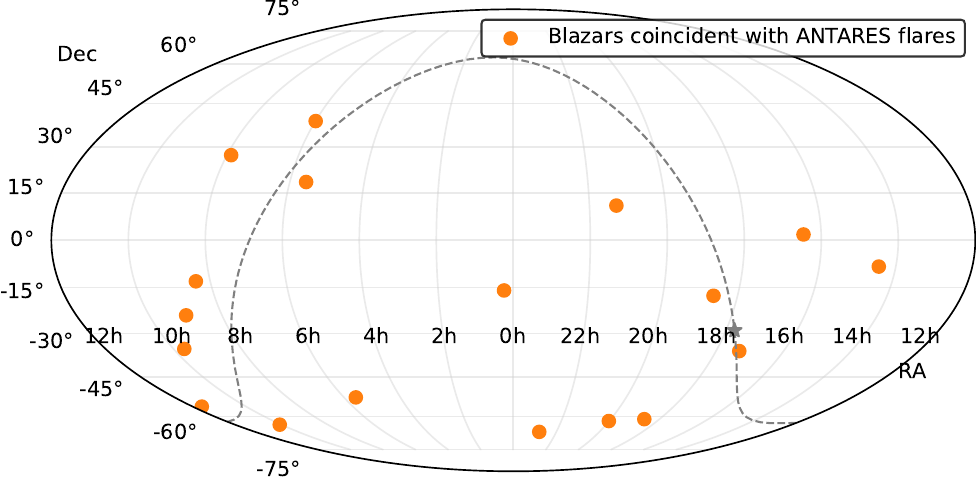}
    \caption{Sky map in equatorial coordinates showing the positions of the VLBI blazars coincident with the $3\sigma$--flares found in ANTARES data (orange dots). See Table \ref{tab:bestsources} for a list of these flares. The location of the Galactic plane (dashed line) and of the Galactic Center (star marker) is also shown.}
    \label{fig:flares_skymap}
\end{figure}

\setlength{\tabcolsep}{.3em}
\begin{table*}[h]

        \centering
    \caption{List of radio-bright blazars for which a pre-trial significance above $3\sigma$ for at least one of the tested time profiles (Gaussian-shaped and box-shaped) has been obtained. The first three columns report the name and equatorial coordinates of the sources. The remaining columns summarise the results of the search in terms of best-fit central time of the flare $\hat{T}_0$, flare duration $\hat{\sigma}_t$, number of signal events $\hat{\mu}_s$, spectral index $\hat{\gamma}$ and pre-trial $p$-value, for the Gaussian-shaped and box-shaped signal time profile. The most significant flare found assuming each of the considered time shape is highlighted in bold.\medskip}
        \label{tab:bestsources}
    
    \resizebox{\linewidth}{!}{
        \begin{minipage}{0.95\textwidth} 
        \centering
 {\def\arraystretch{1.2}
    \footnotesize
    \begin{tabular}{lrr|rrrrl|rrrrl}
      \hline
\multicolumn{3}{c|}{Source} & \multicolumn{10}{c}{Results} \\
   \hline
\multicolumn{1}{c}{Name} & 
\multicolumn{1}{c}{$\delta$} & 
\multicolumn{1}{c|}{$\alpha$} & 
\multicolumn{5}{c|}{\makecell{Gaussian-shaped time profile}} &
\multicolumn{5}{c}{\makecell{Box-shaped time profile}} \\
    &   &  & $\hat{T}_0$ & $\hat{\sigma}_t$  &  $\hat{\mu}_s$ & $\hat{\gamma}$ & $p$-value  & $\hat{T}_0$  & $\hat{\sigma}_t$  &  $\hat{\mu}_s$ & $\hat{\gamma}$ & $p$-value \\
    & [deg] & [deg] & [MJD] & [days] & & & & [MJD] & [days] & & & \\
\hline

J0112$-$6634 & $-$66.6 & 18.1 & 58215 & 304 & 4.5 & 2.7 & 0.0026 & 58154 & 305 & 3.8 & 2.7 & 0.0097 \\

\textbf{J1355$-$6326} &  \textbf{$-$63.4}  &  \textbf{208.9}  &  \textbf{56524}  &  \textbf{1041}  &  \textbf{7.9}  &  \textbf{2.8}  &  \textbf{0.00018}  &  56091  &  905  &  6.0 &  2.9  &  0.0048\\

J0359$-$6154 & $-$61.9 & 59.7 & 56316 & 78 & 5.4 & 3.5 & 0.0022 &  56321 & 112 & 5.7 & 3.5 & 0.0013 \\

J0522$-$6107 & $-$61.1 & 80.6 & 56221 & 42 & 4.6 & 3.4 & 0.0034 & 56232 & 59 & 4.9 & 3.4 & 0.0023 \\

J1220$-$5604 & $-$56.1 & 185.1 &  58406 & 18 & 2.8 & 2.6 & 0.00029 & 58413 & 22 & 0.4 & 2.2 & 0.0032 \\

J1825$-$5230 & $-$52.5 & 276.3 & 57265 & 600 & 5.4 & 2.7 & 0.0031 & 57188 & 959 & 5.5 & 2.7 & 0.0027 \\

J0641$-$3554 & $-$35.9 & 100.3 & 58084 & 16 & 2.9 & 3.0 & 0.0021 & 58081 & 19 & 3.0 & 3.0 & 0.0018 \\

J1418$-$3509 & $-$35.2 & 214.7 & 58120 & 11 & 3.3 & 2.9 & 0.0018 & 58121 & 14 & 2.9 & 2.8 & 0.0021 \\

J1500$-$2358 & $-$24.0 & 225.2 & 55846 & 4 & 3.7 & 2.3 & 0.0016 & 55847 & 6 & 3.7 & 2.2 & 0.0015 \\

J0521$-$1737 & $-$17.6 & 80.3 &  57332 & 1 & 2.0 & 1.9 & 0.0011 & 57333 & 1 & 2.0 & 1.9 & 0.0023 \\ 

J2345$-$1555 & $-$15.9 & 356.3 & 57653 & 460 & 3.2 & 2.6 & 0.0011 & 57784 & 404 & 2.4 & 2.7 & 0.0030 \\

J1537$-$1259 & $-$13.0 & 234.3 & 58201 & 46 & 2.6 & 2.0 & 0.0019 & 58201 & 55 & 2.7 & 2.0 & 0.0016 \\

J0933$-$0819 & $-$8.3 & 143.3 & 57411 & 533 & 3.1 & 2.0 & 0.0014 & 57128 & 697 & 2.9 & 2.0 & 0.0017 \\

J0732+0150 & 1.8 & 113.1 & 55794 & 82 & 5.0 &       3.5 & 0.0010 & 55854 & 61 & 2.7 & 3.4 & 0.033 \\

J0242+1101 & 11.0 & 40.6 & 56676 & 311 & 5.4 & 2.2 & 0.0060 & 56586 & 451 & 6.6 & 2.6 & 0.0021 \\

\textbf{J1826+1831} & \textbf{18.5} & \textbf{276.6} & 57672 & 151 & 2.9 & 2.5 & 0.0015 & \textbf{57636} & \textbf{178} & \textbf{3.0} & \textbf{2.5} & \textbf{0.0010} \\

J1606+2717 & 27.3 & 241.7 &   58793 &         1 & 1.0 & 1.0 & 0.00076 & 58793 & 1 & 1.0 & 1.0 & 0.0017 \\

J1800+3848 & 38.8 & 270.1 & 56590 & 3 & 1.7 & 2.4 & 0.0024 &   56590 & 3 & 1.9 &  2.6 & 0.0021 \\

\hline
\end{tabular}
      }
      \end{minipage}
}
\end{table*}

\begin{figure*}[ht!]
    \includegraphics[width=\textwidth]{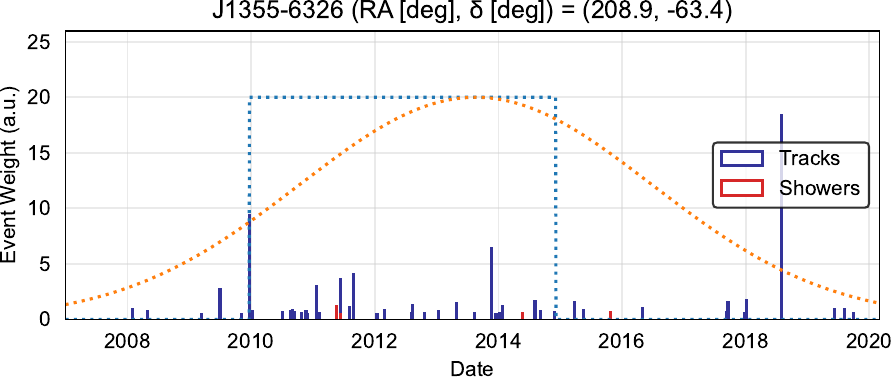}\\
    \vspace{0.00mm}
    \includegraphics[width=\textwidth]{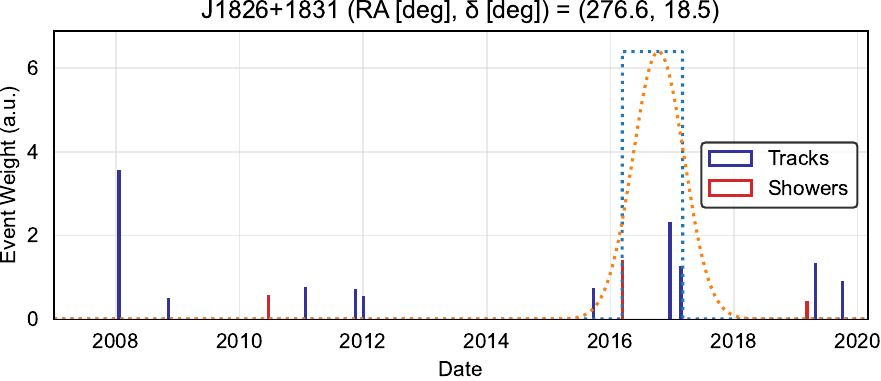}
    \caption{Weighted time distribution of the ANTARES events close to the location of J1355$-$6326 (top) and J1826$+$1831 (bottom). The dotted box and Gaussian time profiles have been drawn using the best-fit values of $\hat{\sigma}_t$ and $\hat{T}_0$ found in each case and reported in Table~\ref{tab:bestsources}. Tracks (showers) are shown in blue (red).}
    \label{fig:TimeDistr1}
\end{figure*}

The search results in 18 sources showing a flare with a pre-trial significance above $3\sigma$ for at least one of the tested time profiles. They are visualized in Figure~\ref{fig:flares_skymap} and listed in Table~\ref{tab:bestsources}, together with the corresponding best-fit values of the free parameters. The most significant Gaussian (box) flare is found from the direction of J1355$-$6326 (correspondingly, J1826$+$1831), with a pre-trial significance of $3.7\sigma$ (correspondingly, $3.3\sigma$) in the two-sided convention. When accounting for the fact that multiple sources have been investigated, a post-trial $p$-value of $29\%$ and $84\%$ is obtained for the best Gaussian and box flares, respectively.
The weighted time distribution of the ANTARES events close to J1355$-$6326 and J1826$+$1831 is shown in Figure~\ref{fig:TimeDistr1}. Only track-like (shower-like) events within a distance of $5^\circ$ ($10^\circ$) from the blazars are included in the plot. A higher weight is associated to events with smaller distance to the source and larger value of the energy estimator.

\begin{figure}[ht]
\centering
    \includegraphics[width=0.6\textwidth]{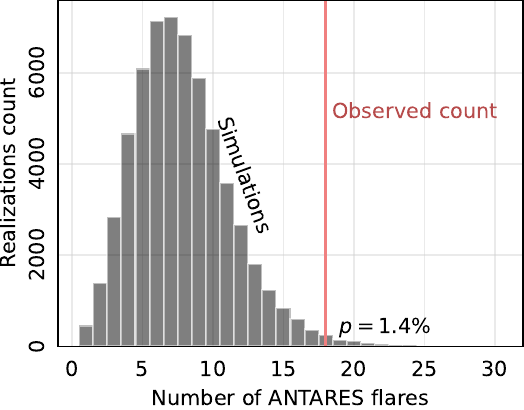}
    \caption{Cumulative excess of $3\sigma$--flares found in ANTARES data. The vertical line indicates the observed count, while the histogram represents the number of sources in random realizations. The probability to observe 18 or more sources with a pre-trial flare significance above $3\sigma$ is $1.4\%$.}
    \label{fig:flares_count}
\end{figure}

While none of the investigated sources alone is observed at a significant detection level when accounting for trials, a further study is performed to check for the presence of a cumulative excess. By performing background PEs in which the same source catalog is targeted, it is found, as shown in Figure~\ref{fig:flares_count}, that the probability to find 18 or more sources with a pre-trial significance greater than $3\sigma$ is $1.4\%$ ($2.5\sigma$). This result provides an additional hint to the time-independent analysis, that a fraction of the blazars contained in the VLBI catalog could be neutrino emitters.

\section{Multi-messenger flares comparison} 
\label{sec:MMflares}

\begin{figure*}[ht!]    \includegraphics[width=\textwidth]{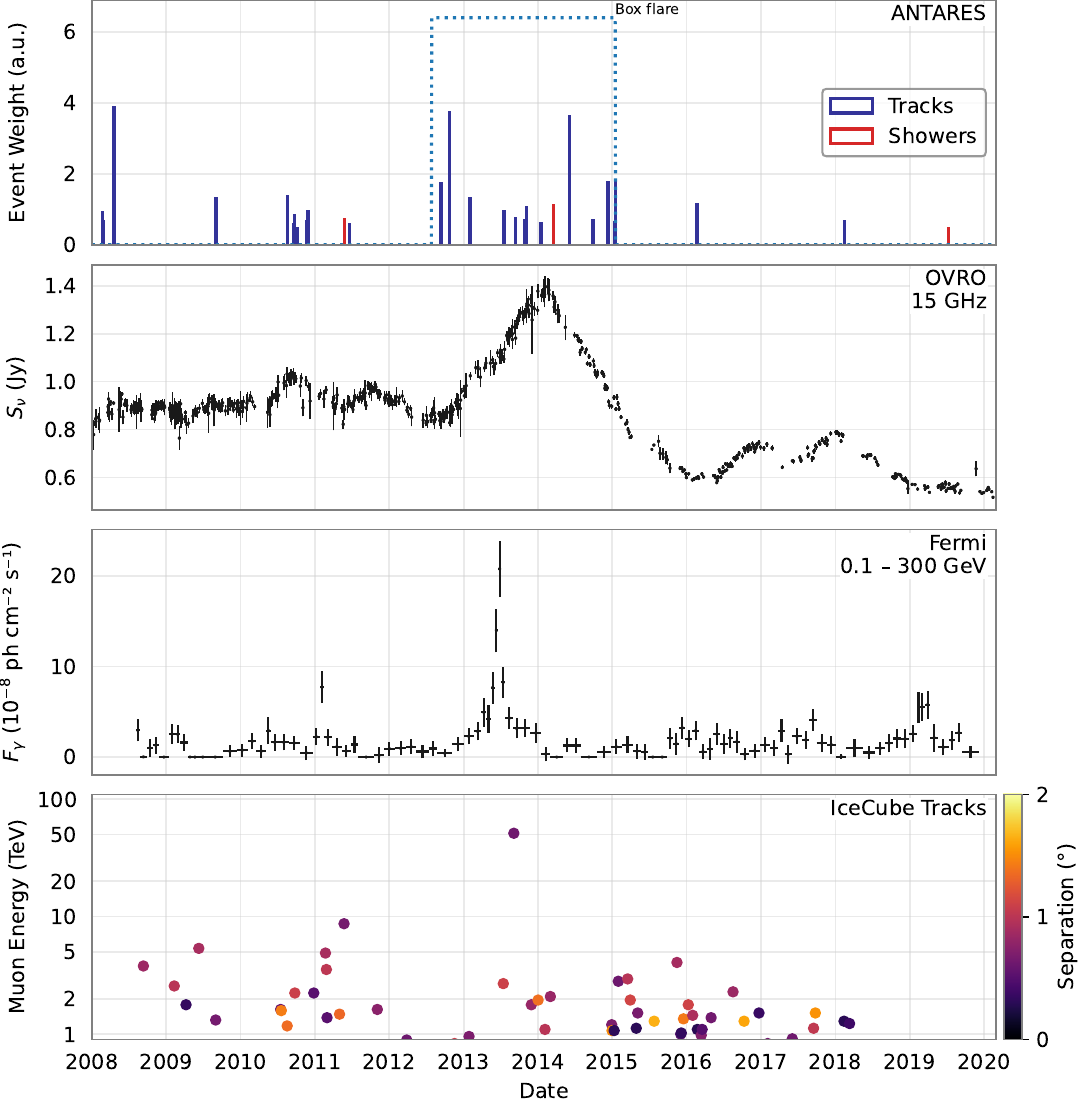}
    
    \caption{Multi-messenger light curves from the direction of the blazar J0242+1101 as a function of time, since 2008. Top panel: weighted time distribution of the ANTARES track-like (shower-like) events within $5^{\circ}$ ($10^{\circ}$) from this object. The box profile has been drawn using the best-fit values of $\hat{\sigma}_t$ and $\hat{T}_0$ found in this analysis. Second panel: OVRO radio light curve. Third panel: adaptive binned $\gamma$-ray light curve obtained from \textit{Fermi} LAT data. Bottom panel: weighted time distribution of the IceCube track-like events closer to J0242+1101 than their $50\%$ angular uncertainty. The applied weight corresponds to the energy of each event. The color scale indicates the event angular distance from the source. }
    \label{fig:flares}
\end{figure*}

As a follow-up study of the findings of this analysis, the obtained best-fit neutrino flares have been compared to the radio light curves produced by the Owens Valley Radio Observatory (OVRO)~\cite{OVRO} for those sources of Table~\ref{tab:bestsources} for which radio data are available. 
A notable overlap in time is noticed between the best-fit neutrino flare found in this analysis from the direction of J0242+1101 (PKS\,0239+108) and its largest flare observed in radio, as shown in Figure~\ref{fig:flares}.
In view of this observation, the time distribution of the public data of the \textit{Fermi} LAT $\gamma$-ray telescope and of the IceCube neutrino telescope compatible with the direction of J0242+1101 have also been studied. Figure~\ref{fig:flares} reports also the adaptive binned $\gamma$-ray light curve, obtained from \textit{Fermi} data using the method described in~\cite{2012A&A...544A...6L,Kramarenko:2021rkf}. Remarkably, the most significant \textit{Fermi} $\gamma$-ray flare for this blazar happened during the flaring emission detected in radio and the period highlighted by the ANTARES analysis. 
Note that the $\gamma$-ray flare peak preceding radio is a typical scenario explained by synchrotron self-absorption of the jet base; this locates the $\gamma$-ray emission region close to the jet base \cite{2010ApJ...722L...7P,Kramarenko:2021rkf}.
The time distribution of the IceCube track-like events in the 10-year point-source sample~\cite{IC10} with direction compatible with the blazar position within the $50\%$ angular error reported by the IceCube Collaboration, is also shown. Only events with an angular uncertainty contour smaller than 10~deg$^2$ are depicted. While there is no evidence of time clustering of the IceCube events, a $\nu_\mu$-induced track with the notable high energy of 50~TeV was detected during the flare, its reported angular uncertainty radius being $1.4^\circ$. Furthemore, it is worth reporting that three IceCube alerts \cite{IceCube:2022ham}, namely 131165:9342044, 129933:32926212, and 128672:38561326, have a likelihood best-fit direction which lies at a distance of $1.4^{^\circ}$, $1.5^{^\circ}$, and $1.6^{^\circ}$, respectively, from J0242+1101. However, their detection date is subsequent to the one of the flare discussed here. 

The neutrino-radio-$\gamma$ flare coincidence in J0242+1101 is an \textit{a-posteriori} non-blind finding. It is still instructive to evaluate how likely such a correlation is to arise by chance. This study is conducted by running the whole flare fitting pipeline on background-only PEs. Each simulation of the analysis results in a fake list of neutrino flares, similar to Table \ref{tab:bestsources}. Then, exactly the same calculations are performed for the real list of flares and for random realizations. Specifically, blazars with the following characteristics are counted:
\begin{itemize}
    \item the global maximum of their radio/$\gamma$ light curve falls within the ANTARES flare duration, 
    $\hat{T}_0 \pm \hat{\sigma}_t$;
    \item that maximum is at least as high as for J0242+1101, compared to the median flux of the same blazar; this corresponds to maximum-to-median ratios above $1.6$ for radio, and above $3.5$ for $\gamma$.
\end{itemize}
Thanks to OVRO and \textit{Fermi} observatories, both radio and $\gamma$-ray light curves are available for hundreds of blazars: the CGRaBS OVRO sample \cite{Hovatta:2020lor} and weekly curves from the \textit{Fermi} LAT light Curve Repository\footnote{\url{https://fermi.gsfc.nasa.gov/ssc/data/access/lat/LightCurveRepository/}}~\cite{Fermi-LAT:2023iml}. There are 335 blazars in the intersection of these samples.

The outcome of this study indicates that it is relatively rare to find even a single matching blazar with the above mentioned characteristics: it appears in only $p=0.5\%$ of random realizations; 4\% for either radio or $\gamma$-rays separately. This fraction of $p=0.5\%$ is the $p$-value, the chance coincidence probability to observe this kind of temporal neutrino-electromagnetic correlation. Given that the J0242+1101 analysis is \textit{a-posteriori}, these findings should be considered as hints to be tested further when more data become available.

\section{Summary} 
\label{sec:conclusions}

A search for associations between 3411 radio-bright blazars of a complete VLBI-flux-density limited all-sky sample, and an ANTARES neutrino event sample has been performed. 

A time-integrated association search was conducted with pair-counting and likelihood-based approaches, using only the track-like neutrino candidates.
The pair-counting method finds an excess of 59 neutrino-blazar pairs, with post-trial $p$-value of $0.03$ ($2.2\sigma$). The likelihood-based method finds a similar $p$-value of $p=0.03$, for an energy spectrum of $E^{-2.3}$ and with the assumption of a neutrino flux proportional to the VLBI flux density. Upper limits on the neutrino flux from VLBI blazars are computed for different values of the spectral index $E^{-\gamma}$, mildly constraining the IceCube HESE measurement around $100$ TeV, where the VLBI blazars would contribute to $\sim 50 \%$ of the flux.  

The time-dependent search for flares, conducted using an unbinned maximum likelihood method and combining track-like and shower-like samples, finds 18 sources with neutrino flares above $3\sigma$ (pre-trial). The two most significant flares come from the blazars J1355$-$6326 ($3.7\sigma$) and J1826$+$1831 ($3.3\sigma$), see Table~\ref{tab:bestsources}. However, neither is significant alone after correcting for multiple trials. A cumulative effect is detected with $p=0.014$ ($2.5\sigma$). This could be a hint of a population of blazars producing neutrinos with strong variations of the flux density.

A notable temporal correspondence between the neutrino flare from the direction of the J0242$+$1101 blazar, and the most prominent wide-band electromagnetic flare of that source is found. The chance probability of such a coincidence between neutrino emission, radio, and $\gamma$-ray light curves is $p=0.5\%$. This is a \textit{post-hoc} estimate that could be a hint for a connection between neutrino and electromagnetic emission. More observational data are required to verify the correlation more robustly and look for other similar coincidences.

The first results on associating ANTARES neutrinos with radio-bright blazars are presented in this paper. Blazar neutrino emission appears highly variable and is likely to correlate with electromagnetic flares. Still, the situation is far from being clear, other classes of neutrino sources should contribute: the neutrino sky can be diverse, including a Galactic component. All neutrino observatories, IceCube, KM3NeT, and Baikal-GVD, are continuously improving to provide more detections with better directional accuracy and reliability. A strengthening of the observed correlation between blazars and neutrinos could then be expected in the coming years. Radio observational programs are ongoing to aid in reliable identification of coincidences and better understanding of associations.

\newpage

\section*{Acknowledgements}

We thank Eduardo Ros for helpful comments on the manuscript.
This research has made use of data from the OVRO 40-m monitoring program, supported by private funding from the California Institute of Technology and the Max Planck Institute for Radio Astronomy, and by NASA grants NNX08AW31G, NNX11A043G, and NNX14AQ89G and NSF grants AST-0808050 and AST-1109911. 

The authors acknowledge the financial support of the funding agencies:
Centre National de la Recherche Scientifique (CNRS), Commissariat \`a
l'\'ener\-gie atomique et aux \'energies alternatives (CEA),
Commission Europ\'eenne (FEDER fund and Marie Curie Program),
LabEx UnivEarthS (ANR-10-LABX-0023 and ANR-18-IDEX-0001),
R\'egion Alsace (contrat CPER), R\'egion Provence-Alpes-C\^ote d'Azur,
D\'e\-par\-tement du Var and Ville de La
Seyne-sur-Mer, France;
Bundesministerium f\"ur Bildung und Forschung
(BMBF), Germany; 
Istituto Nazionale di Fisica Nucleare (INFN), Italy;
Nederlandse organisatie voor Wetenschappelijk Onderzoek (NWO), the Netherlands;
Executive Unit for Financing Higher Education, Research, Development and Innovation (UEFISCDI), Romania;
Grants PID2021-124591NB-C41, -C42, -C43 funded by MCIN/AEI/ 10.13039/501100011033 and, as appropriate, by “ERDF A way of making Europe”, by the “European Union” or by the “European Union NextGenerationEU/PRTR”,  Programa de Planes Complementarios I+D+I (refs. ASFAE/2022/023, ASFAE/2022/014), Programa Prometeo (PROMETEO/2020/019) and GenT (refs. CIDEGENT/2018/034, /2019/043, /2020/049. /2021/23) of the Generalitat Valenciana, EU: MSC program (ref. 101025085), Spain;
Ministry of Higher Education, Scientific Research and Innovation, Morocco, and the Arab Fund for Economic and Social Development, Kuwait.
We also acknowledge the technical support of Ifremer, AIM and Foselev Marine
for the sea operation and the CC-IN2P3 for the computing facilities.
Y.~Y.~K.\ acknowledges support from the M2FINDERS project which has received funding from the European Research Council (ERC) under the European Union’s Horizon 2020 Research and Innovation Programme (grant agreement No.~101018682).
T.~H.\ was supported by the Academy of Finland projects 317383, 320085, 322535, and 345899. 
S.~K.\ acknowledges support from the European Research Council (ERC) under the European Unions Horizon 2020 research and innovation programme under grant agreement No.~771282.

\bibliographystyle{utphys}
\providecommand{\href}[2]{#2}\begingroup\raggedright\endgroup

\end{document}